# Decarbonization patterns of residential building operations in China and India

Ran Yan [1], Nan Zhou [2], Wei Feng [2], Minda Ma [2] [*], Xiwang Xiang [1], Chao Mao [1] [*]

1.  School of Management Science and Real Estate, Chongqing University, Chongqing, 400045, PR China

2.  Building Technology and Urban Systems Division, Energy Technologies Area, Lawrence Berkeley National Laboratory, Berkeley, CA 94720, United States

- The leading corresponding author: Dr. Minda Ma, Email: maminda@lbl.gov
  Homepage: https://buildings.lbl.gov/people/minda-ma

# Graphical abstract

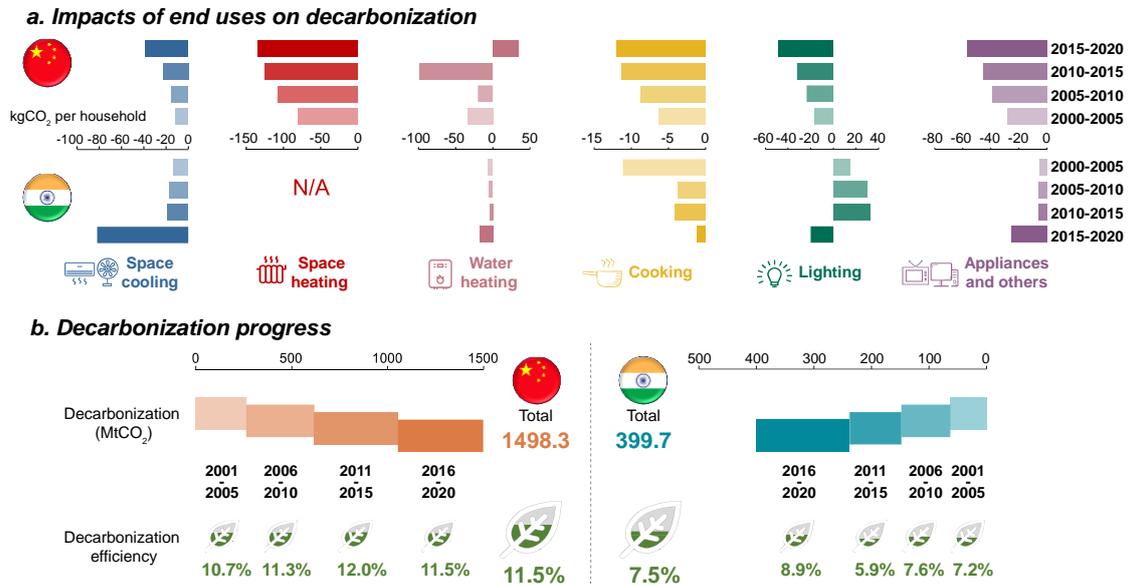

**Graphical abstract.** (a) Impact of six end uses on decarbonization and (b) the decarbonization progress of residential building operations in India and China in 2000-2020.

## Highlights

- Operational carbon intensity in China and India increased by 1.4% and 2.5% per year in 2000-2020;

- Household expenditure-related energy intensity and emission factors were crucial to decarbonization;

- Building electrification significantly promoted the end-uses' emission factor effects on decarbonization;

- China and India decarbonized 1498.3 and 399.7 $MtCO_2$ in residential building operations in 2000-2020;

- For decarbonization intensity, India nearly caught up with China (~180 $kgCO_2$ per household) in 2020.

## Abstract

As the two largest emerging emitters with the highest growth in operational carbon from residential buildings, the historical emission patterns and decarbonization efforts of China and India warrant further exploration. This study aims to be the first to present a carbon intensity model considering end-use performances, assessing the operational decarbonization progress of residential building in India and China over the past two decades using the improved decomposing structural decomposition approach. Results indicate (1) the overall operational carbon intensity increased by 1.4% and 2.5% in China and India, respectively, between 2000 and 2020. Household expenditure-related energy intensity and emission factors were crucial in decarbonizing residential buildings. (2) Building electrification played a significant role in decarbonizing space cooling (-87.7 in China and -130.2 kilograms of carbon dioxide ($kgCO_2$) per household in India) and appliances ($\sim$ -169.7 in China and $\sim$ -43.4 $kgCO_2$ per household in India). (3) China and India collectively decarbonized 1498.3 and 399.7 mega-tons of $CO_2$ in residential building operations, respectively. In terms of decarbonization intensity, India (164.8 $kgCO_2$ per household) nearly caught up with China (182.5 $kgCO_2$ per household) in 2020 and is expected to surpass China in the upcoming years, given the country's robust annual growth rate of 7.3%. Overall, this study provides an effective data-driven tool for investigating the building decarbonization potential in China and India, and offers valuable insights for other emerging economies seeking to decarbonize residential buildings in the forthcoming COP28[a] age.

## Keywords

---

[a] COP28 is the abbreviation of 2023 United Nations Climate Change Conference.

**Abbreviation notation**

DSD – Decomposing structural decomposition

GDIM – Generalized Divisia index method

GDP – Gross domestic product

HCE – Household consumption expenditure

LMDI –Logarithmic mean Divisia index

LPG – Liquified petroleum gas

USD – US dollar

## Nomenclature

$C$ – Carbon emissions

$c$ – Carbon emissions per household (Carbon intensity)

$\mathrm{d}F$ – The slack component introduced in the DSD method

$\mathrm{d}F_i$ ($i = 1, 2, 3, 4, 5, 6$) – The shift component introduced in the DSD method

$\Delta DC$ – Total decarbonization

$\Delta Dc$ – Decarbonization intensity

$\Delta d$ – Decarbonization efficiency

$E$ – Energy consumption

$e$ – Household expenditure-related energy intensity

$g$ – GDP per capita

$H$ – Family households

$k$ – Energy-related carbon intensity of end uses (i.e., the emission factors)

$kgCO_2$ – Kilograms of carbon dioxide

kgce – Kilograms of coal equivalent

$MtCO_2$ – Mega-tons of carbon dioxide

Mtce – Mega-tons of coal equivalent

$m^2$ – Square meters

$P$ – Population size

$p$ – Household size

$s$ – Household expenditure index

$w$ – End-use energy structure

# 1. Introduction

The building sector's operational carbon constitutes 27% of global anthropogenic carbon emissions [1], with the residential sector's share rebounding to a record-high of 60% after COVID-19 restrictions [2]. Developed countries have taken the early lead in urbanization, with large residential building stocks built in the last century [3]. Technological and behavioral lock-in effects significantly limit the operational decarbonization of residential building [4]. As the two largest developing economies actively responding to the global climate crisis [5] and the two emerging emitters with the highest emission increase from residential building [6], the analysis of past emission patterns and mitigation efforts in China and India will inform the formulation of effective climate policies for building and community development. Furthermore, these findings will serve as valuable references for other emerging emitters, contributing to the negotiation of a fair outcome in decarbonization efforts with developed countries [7].

Aside from macroeconomic indicators, discussion on the impact factors of operational carbon emissions from residential buildings should investigate various end-use energy consumption in households [8], directly affected by user behavior and facility technology [9]. Notably, lifestyle changes and technology development in China and India have been extraordinary over the past twenty years [10, 11], along with the continuous advancement of societies and economies. Current studies rarely pursue a comparison of decarbonization changes in residential buildings affected by end-use performances between China and India across the 21$^{st}$ century [12, 13]. Therefore, the following three questions should be addressed to bridge this gap and evaluate the future decarbonization potential of residential buildings in China and India, seeking a fair emissions cap:

- What are the historical processes of operational carbon and the corresponding decarbonization?
- What is the impact of end-use performances on operational decarbonization?
- How can operational decarbonization be further accelerated in China and India?

To address these questions, this study evaluates the operational decarbonization progress of residential buildings in China and India in the 21$^{st}$ century via the improved decomposing structural decomposition (DSD) approach for the first time. Specifically, considering economic, societal, and behavioral aspects, an emission model featuring end-use consumption is developed for identifying factors affecting carbon intensity changes, and the impact of various end uses on the carbon intensity is further investigated. Subsequently, six scales of decarbonization are utilized to evaluate the historical processes of decarbonizing residential buildings. Furthermore, a review and outlook on decarbonization strategies for residential building operations in both India and China are presented to address current challenges and achieve significant decarbonization in the future.

The major novelty is providing a useful tool for different emitters (especially the emerging economies of China and India) to investigate their decarbonization patterns of residential building operations associated with end-use performances over the last two decades. If the major emitters represented by India and China can realize deep decarbonization early, more carbon budget will be released for other emerging economies' development towards the 1.5 °C goal. Thus, it is urgent and necessary to offer an effective data-driven model for evaluating the historical decarbonization patterns in the end-use performances of residential buildings. Especially, the end-use emission model is combined with the improved DSD method to assess the decarbonization potential since this century.

This study is structured as: Section 2 presents the literature review. Section 3 proposes the carbon intensity model associated with end-use performances and the DSD-based decomposition of carbon intensity; data collection is also introduced. Section 4 illustrates the changes and the corresponding drivers of operational carbon intensity, with further analysis of end-uses' impact. In Section 5, Section 5.1 assesses and compares the historical decarbonization of residential buildings in China and India, while Section 5.2 tests the DSD robustness, and Section 5.3 reviews and provides an outlook on the decarbonization strategies for residential buildings in both India and China. In conclusion, Section 6 presents a summary of the findings and future works.

## 2. Literature review

Focusing on the operational carbon of residential building, India and China continue to experience remarkable growth, whereas major advanced economies have displayed stagnant or declining trends over the past two decades [14], as illustrated in Fig. 1. Experience from developed countries suggests that high levels of operational carbon in residential buildings are primarily due to significant carbon lock-in effects [15], with technology lock-in and behavior lock-in being the two main factors [16]. As carriers of energy consumption, various end uses in households pose challenges to decarbonization during the operation stage of residential buildings [17]. With rapid economic growth and improvements in living standards, the carbon lock-in effects of residential building operations in India and China have become increasingly apparent [18].

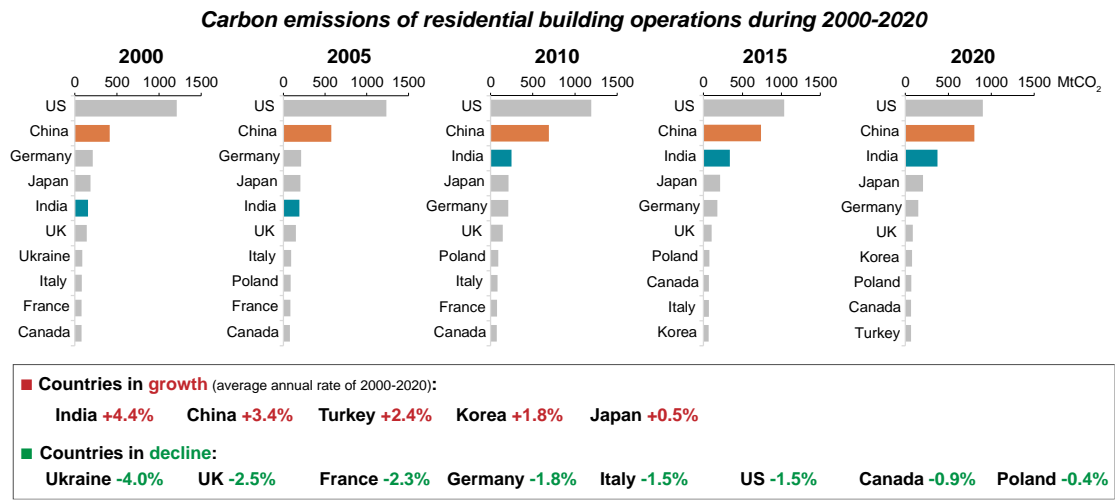

**Fig. 1.** Operational carbon of residential buildings in 2000-2020 among top emitters.

In previous studies, multiple perspectives have been employed to analyze the impact of end uses on energy and emissions of residential building operations in China [19]. For example, Zheng, Wei [12] conducted a household survey to provide insight into the characteristics of end-use energy consumption for a specific year. Some researchers have explored carbon emissions from particular end-use performances in detail (e.g., heating [20], air conditioners [21], and appliances [22]). Swan, et al. [23] reviewed residential energy modeling approaches to further compare end-use energy consumption between China and other countries. In terms of related studies on India, due to the widespread use of dirty energy sources such as biomass and kerosene [24], a low electrification rate, and the slow pace of urbanization [25], researchers have primarily focused on energy consumption rather than carbon emissions when discussing energy transitions in Indian households [13] and energy efficiency improvements for home appliances [26]. Although the India National Sample Survey Organization periodically released detailed information on household energy consumption based on a national household sample survey [27], a time lag and data incoherence have resulted in a scarcity of recent studies related to end-use carbon emissions of residential building operations [28]. Projections of future household energy consumption patterns suggested that India's carbon emissions from residential building operations in 2050 may increase tenfold compared to 2005 levels [29]. Consequently, exploring the decarbonization of end uses is both urgent and significant.

Decomposition methods are employed to quantify the impact of end uses and other crucial factors on emission transition and decarbonization efforts of residential buildings [30], as well as to further evaluate historical decarbonization efforts [31]. Classical decomposition methods include the structure decomposition analysis [32] and the index decomposition analysis such as the logarithmic mean Divisia index (LMDI) [33]. However, these methods have been proven to have limitations. While the structure decomposition analysis allows for a detailed examination of technology and demand changes based on an input–output model [34], it can only work with data from consecutive periods [35], as input–output tables are released every five years rather than annually [36]. To overcome residual interference [37], the LMDI was introduced as an updated version of index decomposition analysis [38]; however, addressing the interdependence of factors remains challenging [39]. In 2014, Vaninsky [40] proposed the generalized Divisia index method (GDIM), which overcomes the abovementioned limitations and has been widely applied in research on indicator decomposition analysis of carbon emissions [41, 42]. Furthermore, Boratyński [43] introduced a simpler and more intuitive decomposition formula based on the GDIM and the Harrison, Horridge, and Pearson decomposition method in 2021, called the DSD method, which has been employed to empirically demonstrate changes in European electricity demand and evaluate the decarbonization of commercial buildings in some top economies [44].

Based on previous studies concerning the assessment and decomposition of carbon emissions, there are two issues that merit further discussion.

**From the end-use perspective, it is worth examining and comparing the decarbonization efforts in residential buildings in the top emerging emitters.** Although there are some related

studies on global or individual scales [45, 46] there is a lack of in-depth analyses focusing on the competition between China and India [47]. Furthermore, due to data acquisition challenges [48], short-term or outdated studies [49] failed to accurately depict the end-use carbon emissions and decarbonization changes in residential building operations influenced by lifestyle changes [50], technology advancements [51], economic recession [52], and the COVID-19 pandemic [53]. A comprehensive comparative analysis at the national level since the millennium is crucial for evaluating the future potential for decarbonization and seeking fair emission allowances.

**Exploring the application of the improved DSD method in characterizing the building carbon intensity is also essential.** In the residential building sector, the conventional GDIM or LMDI methods can't effectively decompose energy intensity or emission factors into individual end uses [54]. However, previous studies have shown that the original DSD method has the potential to cover this gap [44]. Additionally, considering the structural change effects of various end uses will lead to more accurate decomposition results, thereby an extended DSD model is required to achieve this goal.

In response to these issues, this study aims to analyze carbon intensity changes involving the impact of end-use performances using the extended DSD method and evaluate the historical decarbonization efforts of residential building operations in China and India in the first two decades of this century. The primary contributions of this study are as follows:

- **A bottom-up model involving end-use performances is developed to investigate the past emission change of residential building in India and China since the millennium.** Current studies rarely conducted comparative analysis of operational carbon emissions between China and India or examined the long-term effects of end-use performances on operational carbon. The DSD method, previously employed to calculate the operational carbon mitigation in commercial buildings of some economies, is now extended to the residential building sector. By considering economic, societal, and behavioral aspects, a bottom-up model featuring end-use demand is developed for tracking factors influencing carbon intensity changes, combining with the extended DSD method. The impacts of six end-use activities on decarbonizing residential building are further explored, mainly including space cooling, space heating, lighting, cooking, water heating, and appliances.

- **A multiscale assessment with a medium-term time frame is used for the first time to recognize past decarbonization patterns of residential building.** Six scales of decarbonization are proposed to analyze the processes of residential building decarbonization in China and India, encompassing total decarbonization, decarbonization efficiency, decarbonization per household, per capita, per floor area, and per household expenditure. Additionally, a review and outlook of decarbonization strategies for residential buildings in both India and China are presented to address current challenges and achieve deep decarbonization in the future.

## 3. Methods and materials

### 3.1. Emission model of residential building operations

End-use energy consumption ($E$) comprises of primary energy consumption, retail electricity and heating, excluding energy systems losses. The operational energy consumption in family households is carried out by various end uses. Due to the differences in energy structure, geography, climate, and lifestyle between China and India, this study decomposed the end uses of residential building into six types: space cooling, space heating, lighting, water heating, cooking, and appliances with others. To further analyze the effects of end uses on the mitigation of residential buildings between China and India, carbon emissions were generated by these six end-use activities, which can be expressed in Eq. (1). For the sake of data acquisition and comparative analysis, this study mainly discussed the carbon emissions released by energy commodities, excluding biofuels and waste.

$$C = C_{\text{space cooling}} + C_{\text{space heating}} + C_{\text{water heating}} + C_{\text{cooking}} + C_{\text{lighting}} + C_{\text{appliances and others}}$$
$$\text{Shorten as } C = \sum_{i=1}^{6} C_i \tag{1}$$

Considering the aspects of economy, society, and behavior, the carbon emission intensity, which refers to carbon emissions per household, is influenced by six factors: population, households, gross domestic product (GDP), household consumption expenditure (HCE), energy consumption, and carbon emissions. These factors are interrelated and can be represented by the following identity, with abbreviations explained in the Nomenclature:

$$c = \frac{C}{H} = \frac{E}{HCE} \cdot \frac{C}{E} \cdot \frac{P}{H} \cdot \frac{GDP}{P} \cdot \frac{HCE}{GDP} \tag{2}$$

Combining Eqs. (1)-(2), the carbon emission intensity associated with end use $i$ in residential building operations can be written as:

$$c_i = \frac{C_i}{H} = \frac{E_i}{HCE} \cdot \frac{C_i}{E_i} \cdot \frac{P}{H} \cdot \frac{GDP}{P} \cdot \frac{HCE}{GDP} \tag{3}$$

For intuitive expression, $c_i = \frac{C_i}{H}$ represents the carbon intensity associated with end use $i$ in residential building operations, $p = \frac{P}{H}$ represents the population per household (household size), $g = \frac{GDP}{P}$ represents GDP per capita, $s = \frac{HCE}{GDP}$ represents the household expenditure index, $e_i = \frac{E_i}{HCE}$ represents the household expenditure-related energy intensity of end use $i$, $k_i = \frac{C_i}{E_i}$ represents the energy-related carbon intensity (AKA the emission factor) of end use $i$. Thus, Eq. (3) can be simplified as follows:

$$c_i = e_i \cdot k_i \cdot p \cdot g \cdot s$$
$$\text{Summed up as } c = \sum_{i=1}^{6} c_i \tag{4}$$

### 3.2. DSD-based decomposition of carbon intensity

The carbon intensity of residential building operations can vary depending on time and location due to factors such as energy transition, population growth, and economic growth. To better understand the contribution of changes in the end-use energy structure to overall carbon intensity, this study utilized the DSD method, a simpler and more direct decomposition method based on the framework of the GDIM. By applying the DSD approach to the emission model described in Section 3.1, this study was able to decompose the operational carbon intensity of residential building in India and China from 2000 to 2020 and explore the various factors contributing to carbon intensity changes.

The DSD method involves introducing $e$ as the sum of the household expenditure-related energy intensity, which is defined as $e = \sum_{j=1}^{6} e_i$. Additionally, $w_i$ is defined as the share of $E_i$ in $E$, or end-use energy structure, which can also be expressed as $w_i = \frac{E_i}{E}$. For the equation $e_i = \frac{E_i}{HCE}$ in Section 3.1, it is also true that $w_i = \frac{e_i}{e}$. Eq. (4) was established as:

$$c = \sum_{i=1}^{6} e \cdot k_i \cdot w_i \cdot p \cdot g \cdot s \tag{5}$$

Then, Eq. (6) is required to derive the total differential as follows:

$$\mathrm{d}c = \sum_{i=1}^{6} \left( \frac{\partial c_i}{\partial e} \mathrm{d}e + \frac{\partial c_i}{\partial p} \mathrm{d}p + \frac{\partial c_i}{\partial g} \mathrm{d}g + \frac{\partial c_i}{\partial s} \mathrm{d}s + \frac{\partial c_i}{\partial k_i} \mathrm{d}k_i + \frac{\partial c_i}{\partial w_i} \mathrm{d}w_i \right) \tag{6}$$

Moreover, the shift component $\mathrm{d}F_i$ and slack component $\mathrm{d}F$ are introduced to express the change in $w_i$. The expanded linear equations can be yielded in Eq. (7):

$$\begin{cases} \mathrm{D}c = \sum_{i=1}^{6} \left( \frac{\partial c_i}{\partial e} \mathrm{d}e + \frac{\partial c_i}{\partial p} \mathrm{d}p + \frac{\partial c_i}{\partial g} \mathrm{d}g + \frac{\partial c_i}{\partial s} \mathrm{d}s + \frac{\partial c_i}{\partial k_i} \mathrm{d}k_i + \frac{\partial c_i}{\partial w_i} \mathrm{d}w_i \right) \\ \mathrm{d}w_i = \mathrm{d}F_i + \mathrm{d}F \\ \sum_{i=1}^{6} \mathrm{d}w_i = 0 \end{cases} \tag{7}$$

Rewriting Eq. (8) in matrix form is as follows:

$$
\begin{bmatrix}
1 & -\frac{\partial c_1}{\partial w_1} & -\frac{\partial c_2}{\partial w_2} & -\frac{\partial c_3}{\partial w_3} & -\frac{\partial c_4}{\partial w_4} & -\frac{\partial c_5}{\partial w_5} & -\frac{\partial c_6}{\partial w_6} & 0 \\
0 & 1 & 0 & 0 & 0 & 0 & 0 & -1 \\
0 & 0 & 1 & 0 & 0 & 0 & 0 & -1 \\
0 & 0 & 0 & 1 & 0 & 0 & 0 & -1 \\
0 & 0 & 0 & 0 & 1 & 0 & 0 & -1 \\
0 & 0 & 0 & 0 & 0 & 1 & 0 & -1 \\
0 & 0 & 0 & 0 & 0 & 0 & 1 & -1 \\
0 & 1 & 1 & 1 & 1 & 1 & 1 & 0
\end{bmatrix}
\cdot
\begin{bmatrix}
dc \\ dw_1 \\ dw_2 \\ dw_3 \\ dw_4 \\ dw_5 \\ dw_6 \\ dF
\end{bmatrix}
=
$$

$$
\begin{bmatrix}
\sum\frac{\partial c_i}{\partial e} & \sum\frac{\partial c_i}{\partial p} & \sum\frac{\partial c_i}{\partial g} & \sum\frac{\partial c_i}{\partial s} & \frac{\partial c_1}{\partial k_1} & \frac{\partial c_2}{\partial k_2} & \frac{\partial c_3}{\partial k_3} & \frac{\partial c_4}{\partial k_4} & \frac{\partial c_5}{\partial k_5} & \frac{\partial c_6}{\partial k_6} & 0 & 0 & 0 & 0 & 0 & 0 \\
0 & 0 & 0 & 0 & 0 & 0 & 0 & 0 & 0 & 0 & 1 & 0 & 0 & 0 & 0 & 0 \\
0 & 0 & 0 & 0 & 0 & 0 & 0 & 0 & 0 & 0 & 0 & 1 & 0 & 0 & 0 & 0 \\
0 & 0 & 0 & 0 & 0 & 0 & 0 & 0 & 0 & 0 & 0 & 0 & 1 & 0 & 0 & 0 \\
0 & 0 & 0 & 0 & 0 & 0 & 0 & 0 & 0 & 0 & 0 & 0 & 0 & 1 & 0 & 0 \\
0 & 0 & 0 & 0 & 0 & 0 & 0 & 0 & 0 & 0 & 0 & 0 & 0 & 0 & 1 & 0 \\
0 & 0 & 0 & 0 & 0 & 0 & 0 & 0 & 0 & 0 & 0 & 0 & 0 & 0 & 0 & 1 \\
0 & 0 & 0 & 0 & 0 & 0 & 0 & 0 & 0 & 0 & 0 & 0 & 0 & 0 & 0 & 0
\end{bmatrix}
\cdot
\begin{bmatrix}
de \\ dp \\ dg \\ ds \\ dk_1 \\ dk_2 \\ dk_3 \\ dk_4 \\ dk_5 \\ dk_6 \\ dF_1 \\ dF_2 \\ dF_3 \\ dF_4 \\ dF_5 \\ dF_6
\end{bmatrix}
\tag{8}
$$

Alternatively, it can be abbreviated in a general notation as follows:

$$\mathbf{A} \cdot \mathrm{d}\mathbf{y} = \mathbf{B} \cdot \mathrm{d}\mathbf{z} \tag{9}$$

where $\mathbf{A}$ and $\mathbf{B}$ are both the coefficient matrices associated with the endogenous variables $\boldsymbol{y}$ and the exogenous variables $\mathbf{z}$, that is, $\mathbf{A} = f(\mathbf{z}, \mathbf{y})$ and $\mathbf{B} = f(\mathbf{z}, \mathbf{y})$.

Eq. (9) can be further expanded to:

$$\mathbf{A} \cdot \mathbf{dy} = \mathbf{B} \cdot diag(\mathrm{d}\mathbf{z}) \cdot \mathbf{j} \tag{10}$$

where $diag(\mathbf{dz})$ is the diagonal matrix made up of the differential element $\mathbf{dy}$, and $\mathbf{j}$ is a vector consisting of 1. Thus, $\mathbf{dy}$ can be solved in Eq. (11):

$$\mathbf{dy} = \mathbf{A}^{-1} \cdot \mathbf{B} \cdot diag(d\mathbf{z}) \cdot \mathbf{j} \tag{11}$$

It should be mentioned that the core DSD methodology is rooted in Euler's numerical integration method. This means that the changes in exogenous variables are divided into numerous, equally sized segments that are not infinitesimal but still very small. By doing so, the cumulative impact of exogenous variables on endogenous variables was estimated more approximately and more exactly by using the total differential.

Based on the approach described above, let $\mathrm{d}\mathbf{z} = \frac{\Delta z}{N}$, where $N$ represents the number of segments, and $\mathbf{dz}$ is a vector of non-infinitesimal but very small changes in exogenous variables.

According to the original research article on the DSD method, this study set $N = 16000$ to ensure sufficient accuracy. For each segment of a value of $n$ ($n = 1, 2, \cdots, N$), the effect can be represented mathematically as follows:

$$\begin{cases} \mathbf{D}^{(n)} = \left(\mathbf{A}^{(n-1)}\right)^{-1} \cdot \mathbf{B}^{(n-1)} \cdot diag(d\mathbf{z}) \\ d\mathbf{y}^{(n)} = \mathbf{D}^{(n)} \cdot \mathbf{j} \\ \mathbf{z}^{(n)} = \mathbf{z}^{(n-1)} + d\mathbf{z} \\ \mathbf{y}^{(n)} = \mathbf{y}^{(n-1)} + d\mathbf{y}^{(n)} \\ \mathbf{A}^{(n)} = f\left(\mathbf{z}^{(n)}, \mathbf{y}^{(n)}\right) \\ \mathbf{B}^{(n)} = g\left(\mathbf{z}^{(n)}, \mathbf{y}^{(n)}\right) \end{cases} \tag{12}$$

In the initial period, which is reflected as $n=0$, the coefficient matrices $\mathbf{A}$ and $\mathbf{B}$ are initialized as $\mathbf{A}^{(0)}$ and $\mathbf{B}^{(0)}$, respectively. The desired decomposition is obtained by repeatedly computing each iteration and summing the contributions:

$$\begin{cases} \mathbf{D} = \sum\nolimits_{n=1}^{N} \mathbf{D}^{(n)} \end{cases} \tag{13}$$

In this study, the method described above can be used to measure the impact of changes in exogenous variables on carbon intensity. Using the DSD method, the changes in carbon intensity can be decomposed as:

$$\Delta c|_{0 \to T} = \Delta e_{\text{DSD}} + \Delta p_{\text{DSD}} + \Delta g_{\text{DSD}} + \Delta s_{\text{DSD}} + \Delta k_{\text{DSD}} + \Delta w_{\text{DSD}} \tag{14}$$

where $\Delta k_{\text{DSD}}$ and $\Delta w_{\text{DSD}}$ can be further decomposed based on Eq. (5), which yields the following:

$$\Delta k_{\text{DSD}} = \sum\nolimits_{i=1}^{6} \Delta k_i \tag{15}$$

and:

$$\Delta w_{\text{DSD}} = \sum\nolimits_{i=1}^{6} \Delta w_i \tag{16}$$

Eq. (15) reveals the impact of emission factors from various end uses on emission intensity, while Eq. (16) reveals how changes in end-use energy structure affect the intensity. Together with DSD-based decomposition, the factors affecting the operational carbon intensity of residential building are depicted in Fig. 2.

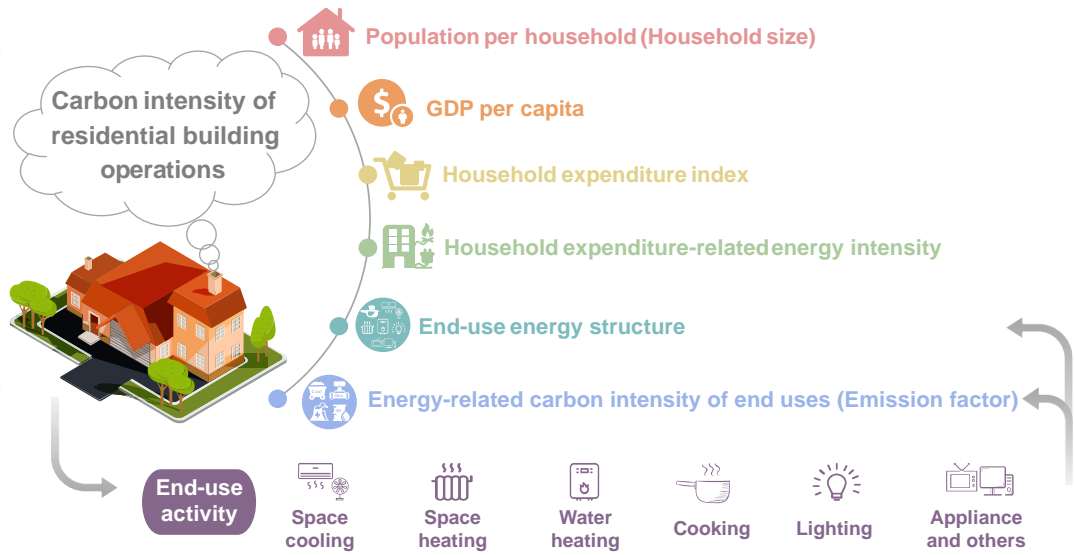

**Fig. 2.** Factor identification of the operational carbon intensity of residential building.

### 3.3. Variables and data sources

The details of the variables involved in Sections 3.1 and 3.2 are refined in Table 1.

**Table 1.** Variable interpretation.

| Variable | Definition | Unit | Expression |
|---|---|---|---|
| $C$ | Carbon emissions | Mega-tons of carbon dioxide (MtCO$_2$) | – |
| $E$ | Energy consumption | Mega-tons of coal equivalent (Mtce) | – |
| $P$ | Population size | Million persons | – |
| $H$ | Family households | Million households | – |
| $GDP$ | Gross domestic product | Million US dollar (USD) | – |
| $HCE$ | Household consumption expenditure | Million USD | – |
| $c$ | Carbon emissions per household (Carbon intensity) | kilograms of carbon dioxide (kgCO$_2$) per household | $c = \dfrac{C}{H}$ |
| $p$ | Household size | Persons per household | $p = \dfrac{P}{H}$ |
| $g$ | GDP per capita | USD/person | $g = \dfrac{GDP}{P}$ |
| $s$ | Household expenditure index | % | $s = \dfrac{HCE}{GDP}$ |
| $e$ | Household expenditure-related energy intensity | kilograms of coal equivalent (kgce)/USD | $e = \dfrac{E}{HCE}$ |
| $k$ | Energy-related carbon intensity of end uses (Emission factors) | kgCO$_2$/kgce | $k = \dfrac{C}{E}$ |
| $w$ | End-use energy structure | % | $w_i = \dfrac{e_i}{e}$ |

Data for this study were collected through various sources. The carbon emissions of residential building operations were obtained from the International Building Emissions Dataset (IBED, https://ibed.world), which used the International Energy Agency dataset (https://www.iea.org/) as the primary benchmark to compile comprehensive and trustworthy data tables for each emitter. For India and China, the building energy and the corresponding emissions data were calculated and calibrated based on the local benchmark datasets (China: http://www.cbeed.cn; India: https://www.india.gov.in/nsso-reports-publications). Especially, IBED covers energy and emissions of various end uses from 2000 to 2020. Demographic and economic indicators of India and China were derived from the World Bank (data.worldbank.org). These sources provided information on population size, family households, GDP, and HCE.

## 4. Results

### 4.1. Carbon intensity changes in residential building operations

Fig. 3 illustrates the changes in operational carbon intensity generated by residential buildings in India and China during 2000-2020, as assessed by the DSD approach. It is evident that the overall carbon intensity has increased for both countries over the last two decades, with a higher average growth rate of 2.5% per year in India and 1.4% per year in China. By dividing the decades into four stages of five years each, the performance of each stage between China and India dynamically changed according to the technical, socioeconomic, and related strategies.

In China, the operational carbon intensity experienced a gradual slowdown in growth rate between 2000 and 2010, reaching an annual peak in 2012 at 1606.4 kgCO$_2$ per household. Since 2010, the Chinese government has further strengthened its decarbonization strategies. Consequently, the operational carbon intensity changes manifested as continuous and slight declines between 2010 and 2020, with an average of -0.9% per year (from 2013 to 2020).

In India, the operational carbon intensity of residential buildings demonstrated sustained growth between 2000 and 2015, with the growth rate initially accelerating before slowing down to approximately zero. From 2010 to 2015, the carbon intensity growth rate reached 24.7%, which then rapidly decreased to -0.1% from 2015 to 2020. Notably, the operational carbon intensity in India reached its annual peak at 1283.3 kgCO$_2$ per household in 2018, before decreasing by -2.6% each year in 2019 and 2020. This decline in operational carbon intensity was due to India's economic contraction amid the COVID-19 pandemic, resulting in restrictions on fossil fuel consumption.

Further analysis of the factors affecting changes in carbon intensity revealed that GDP per capita was the most important trigger. It was also the only one of the six factors consistently demonstrating a positive contribution. From 2000 to 2020, the total contribution of GDP per capita in China was 226.4%, while in India, it was 144.5%. Regarding the negative contributors that reduced carbon intensity, household expenditure-related energy intensity was the most significant negative contributor in both China (-90.6%) and India (-78.1%). This was followed by emission factors, which contributed a negative effect of -87.4% in China and -23.0% in India. Household size also had a negative impact, with modest effects in both China (-35.7%) and India (-13.2%). Furthermore, the influence of the household expenditure index on carbon intensity changes in India and China were unstable, displaying similar trends; that is, it contributed a negative effect before 2010 and a positive effect after 2010. Additionally, the end-use energy structure had minor impact on changes in carbon emission intensity, which will be discussed in detail in Section 4.2. Overall, the analysis of carbon intensity changes in residential buildings in China and India partially answers Question 1, which is raised in Section 1.

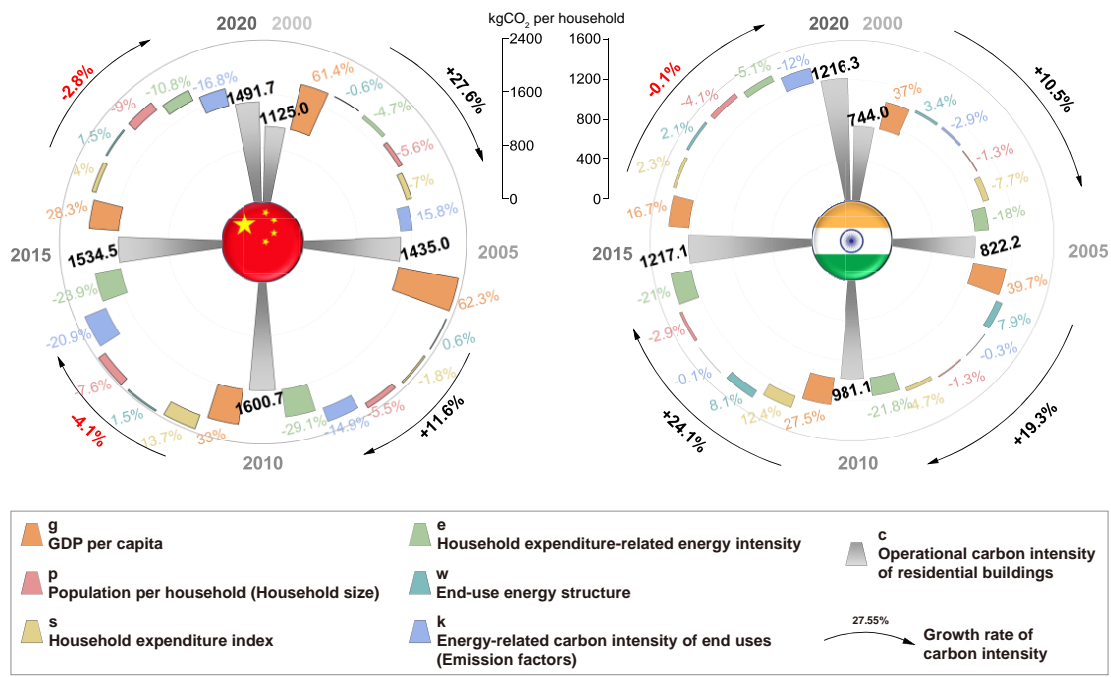

**Fig. 3.** Operational carbon intensity changes in residential buildings in China and India (2000–2020). Note: The period was divided into four stages: 2000-2005, 2005-2010, 2010-2015, and 2015-2020; the operational carbon intensity changes were decomposed into the six factors' impact in each stage.

### 4.2. Impact of end uses on the operational carbon intensity changes

This study also investigated the influence of different end-use performances on the operational carbon intensity. Specifically, the contribution of emission factors (characterized as pale blue fans in Fig. 3) was divided into the effects of six end uses (as shown in Fig. 4). Then, the fan charts in Fig. 4 a and c present the decomposition of emission factor contribution rates across different end uses at four stages from 2000 to 2020 for India and China, respectively. The bar charts in Fig. 4 b transform the contribution rate into the absolute value of carbon intensity changes for a more intuitive investigation.

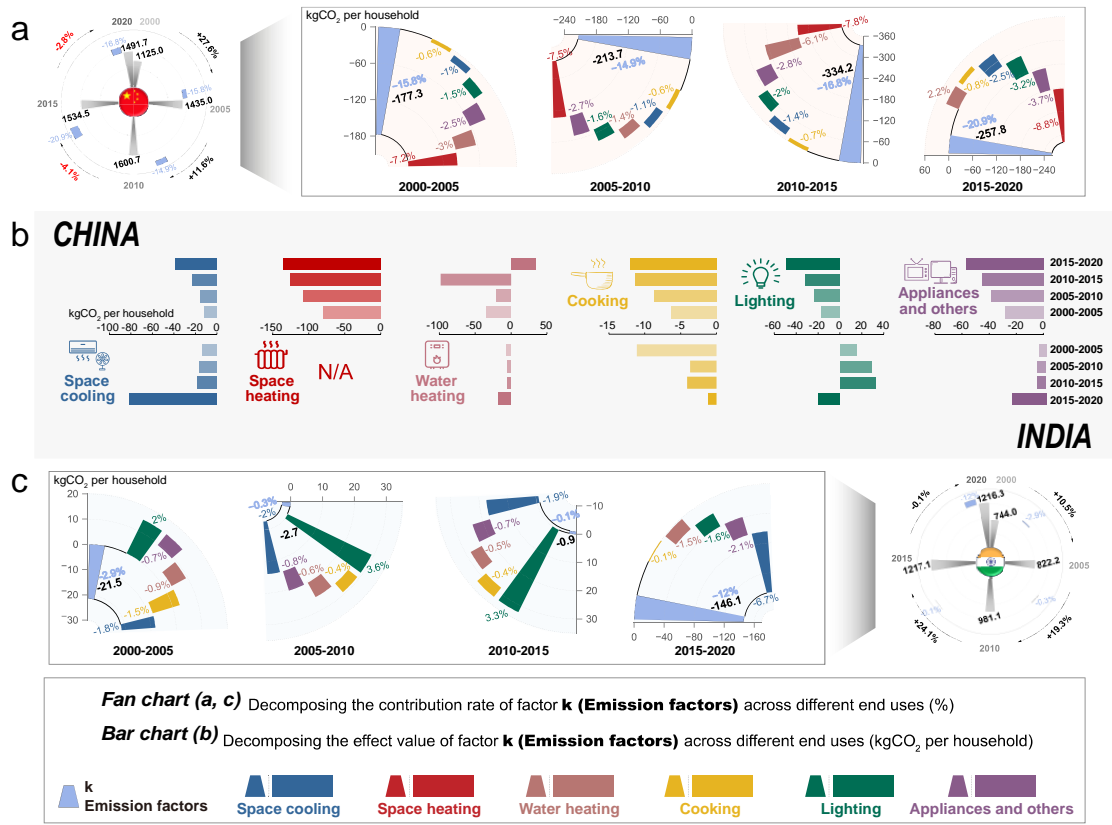

**Fig. 4.** Emission factor effects of various end uses on carbon intensity changes in residential buildings in (a) China and (c) India from 2000 to 2020; (b) absolute value of the emission factor effects on carbon intensity changes among various end uses (2000-2020). Note: Space heating was not involved in India, as the nation locates in a tropical climate zone.

In general, the emission factor effect of each end use had a negative impact on the operational carbon intensity in China; meaning it made a positive contribution to the decarbonization in China. Throughout the four stages from 2000 to 2020, the negative effects of the six contributors on carbon intensity exhibited gradual growth. This can be attributed to China's evolving energy structure over the past twenty years, which has improved the electrification of residential building operations (i.e., the electrification rate, presented by the share of electricity consumption in the total energy demand [55], increased from 4.5% in 2000 to 27.2% in 2020) and the growing popularity of natural gas in urban areas (i.e., the proportion of users rose from 1.8% in 2000 to 29.3% in 2020).

However, unlike China, not all emission factor effect of each end use had a negative impact on the operational carbon intensity in India. The emission factor effects of space cooling, appliances and others maintained a negative contribution from 2000 to 2020, while lighting shifted from a positive contribution to a negative contribution to carbon intensity between 2015 and 2020. The emission factor effects of end uses related to electricity consumption promoted the negative contribution to the carbon intensity in India, particularly between 2015 and 2020. This situation confirms the improvement of the electrification rate in India's residential building operations, which increased from 5.4% in 2000 to 17.8% in 2020.

Specifically, for each end use in China, the most significant negative contributor was space heating, with a total contribution rate of -39.8% and an overall impact on carbon intensity change of -447.9 kgCO$_2$ per household. This was followed by appliances and others, with a total contribution rate of -15.1% (-169.7 kgCO$_2$ per household). Furthermore, lighting, water heating, space cooling, and cooking contributed -10.7% (-120.6 kgCO$_2$ per household), -10.6% (-118.8 kgCO$_2$ per household), -7.8% (-87.7 kgCO$_2$ per household), and -3.4% (-38.4 kgCO$_2$ per household), respectively. In India, the most significant negative contributor was space cooling, with a total contribution rate of -17.5% and an overall impact on carbon intensity changes of -130.2 kgCO$_2$ per household from 2000 to 2020. This was followed by appliances and others, with a total contribution rate of -5.8% (-43.4 kgCO$_2$ per household). In contrast, while lighting turned into a negative effect between 2015 and 2020, it remained the most positive contributor, with a total contribution rate of 7.6% (56.4 kgCO$_2$ per household). Water heating and cooking contributed -4.6% (-33.9 kgCO$_2$ per household) and -2.7% (-20.2 kgCO$_2$ per household), respectively. Space heating in India was not considered in this study, as mainland India is primarily located in the tropics, with the coldest winter temperature being ~ 16 °C. Although the need for space heating exists in the cold Himalayan region, low-income families typically use biomass such as firewood instead of energy commodities for heating. It can be argued that there is minimal demand for energy commodities for space heating in India.

Please note that the emission factor effect of water heating was unstable from year to year in both China and India, mainly due to the multiple energy sources for water heating in poorer towns and rural areas. The emission factor effect of water heating exhibited a slight positive contribution between 2015 and 2020 in China, as capacity and power limitation led an increasing number of rural households to choose gas water heaters over solar or electric options. In India, this unstable contribution is more pronounced, with the contribution rates of water heating and cooking fluctuating each year due to a wide variety of household fuels and large-scale energy consumption of biofuels and waste not involved in this work.

Overall, the emission factor effects of end uses efficiently drove the residential building decarbonization in the top emerging economies, namely China and India. Electrification played a crucial catalytic role in promoting the contribution of space cooling, appliances with others, and lighting to decarbonization. Additionally, energy optimization of residential space heating in China proved fruitful for decarbonization. However, household energy consumption for water heating and cooking in underdeveloped areas has yet to be systematized and purified, posing challenges the operational decarbonization in both India and China. Furthermore, changes in the end-use energy structure also affected the operational carbon intensity changes in residential buildings, resulting in the structural change effects revealed by the DSD method due to the interdependence of individual shares. Since the structural change effects contributed minimally to carbon intensity changes (see Fig. 5), a detailed analysis of this part is not included in this study. In conclusion, the results presented above illustrate the impact of different end uses on the operational decarbonization of residential buildings in China and India, providing an answer to Question 2 raised in Section 1.

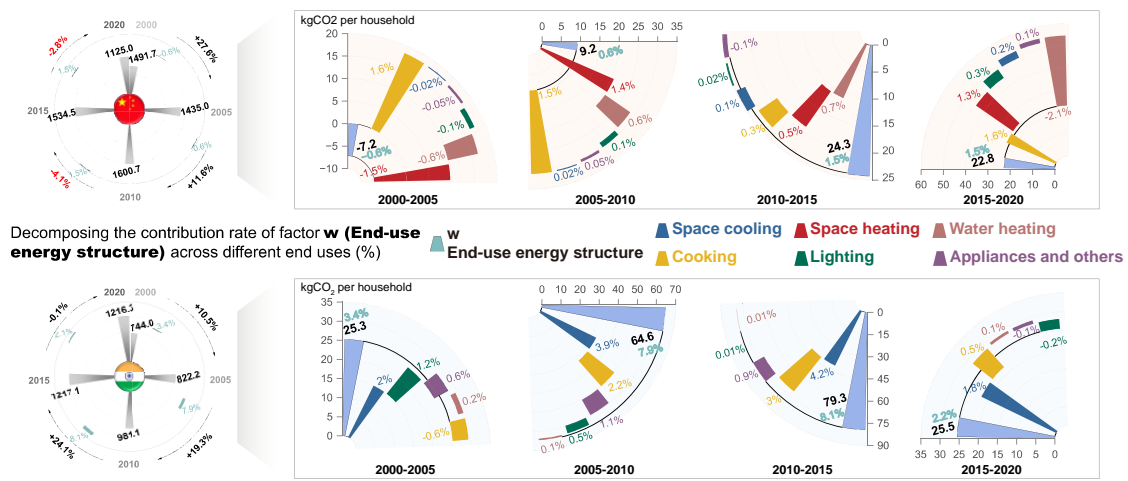

**Fig. 5.** Structural change effects of various end uses on carbon intensity changes in residential building operations in China and India from 2000 to 2020.

# 5. Discussion

## 5.1. Historical decarbonization of residential building operation

Recent United Nations climate actions delimited carbon rights based on national responsibility. However, in regards to the decarbonization of residential building operations, this study argued that assessing the historical performance and future potential of decarbonization solely based on the total amount of decarbonization is insufficient due to differences in territory, population, and economy [56]. Consequently, six scales of decarbonization were employed to comprehensively track the past decarbonization of residential building: total decarbonization, decarbonization efficiency, decarbonization intensity (i.e., decarbonization per household), decarbonization per capita, decarbonization per floor area, and decarbonization per household expenditure.

As stated by the intensity decomposition of carbon emissions, the decarbonization intensity $\Delta Dc$ of residential building operations can be calculated as the sum of the factors' negative contribution in period $\Delta T$ of Eq. (14), as shown:

$$\Delta Dc|_{0 \to T} = -\sum \left( \Delta c(X_j)|_{0 \to T} \right)$$

Where $\Delta c(X_j) \in (\Delta e_{DSD}, \Delta p_{DSD}, \Delta g_{DSD}, \Delta s_{DSD}, \Delta k_{DSD}, \Delta w_{DSD}), \Delta c(X_j)|_{0 \to T} < 0$ (17)

Consequently, the total decarbonization ($\Delta DC|_{0 \to T}$) and the decarbonization efficiency ($\Delta d|_{0 \to T}$) can be expressed in Eqs. (18) - (19):

$$\Delta DC|_{0 \to T} = H|_{0 \to T} \times (\Delta Dc|_{0 \to T})$$ (18)

$$\Delta d|_{0 \to T} = \frac{\Delta DC|_{0 \to T}}{C|_{0 \to T}}$$ (19)

The decarbonization at the scales of per capita, per floor area, and per household expenditure can also be calculated using the above formula. Fig. 6 illustrates the development of total decarbonization and decarbonization efficiency in China and India from 2000 to 2020. From China's perspective, the total cumulative decarbonization reached 1498.3 MtCO$_2$, with a decarbonization efficiency of 11.5% over the past 20 years. During the four stages, the shares of phased accumulation in total cumulative decarbonization were 17% (2001-2005), 24% (2006-2010), 29% (2011-2015), and 30% (2016-2020). The decarbonization efficiency of each stage varied slightly within the range of 11%-12%. These facts reflected the steady improvement of decarbonization in Chinese residential building, with an average increase of 5.7% per year. In comparison, India's total cumulative decarbonization amounted to 399.7 MtCO$_2$, with a corresponding decarbonization efficiency of 7.5%. Moreover, the shares of phased accumulation in total cumulative decarbonization were 16% (2001-2005), 21% (2006-2010), 23% (2011-2015), and 40% (2016-2020), indicating robust growth in decarbonization since 2015, with an average increase of 7.3% per year. The decarbonization efficiency of each stage also showed an upward trend, even reaching a high value of 13.6% in 2020. In conclusion, neither China nor India have reached a significant annual peak of decarbonization for residential building operations, and there is still room for further improvement. On the other hand, due to the gradual slowdown of the decarbonization process in developed countries over recent years, large developing countries like China and India will face increasing pressure to meet the 1.5 °C target.

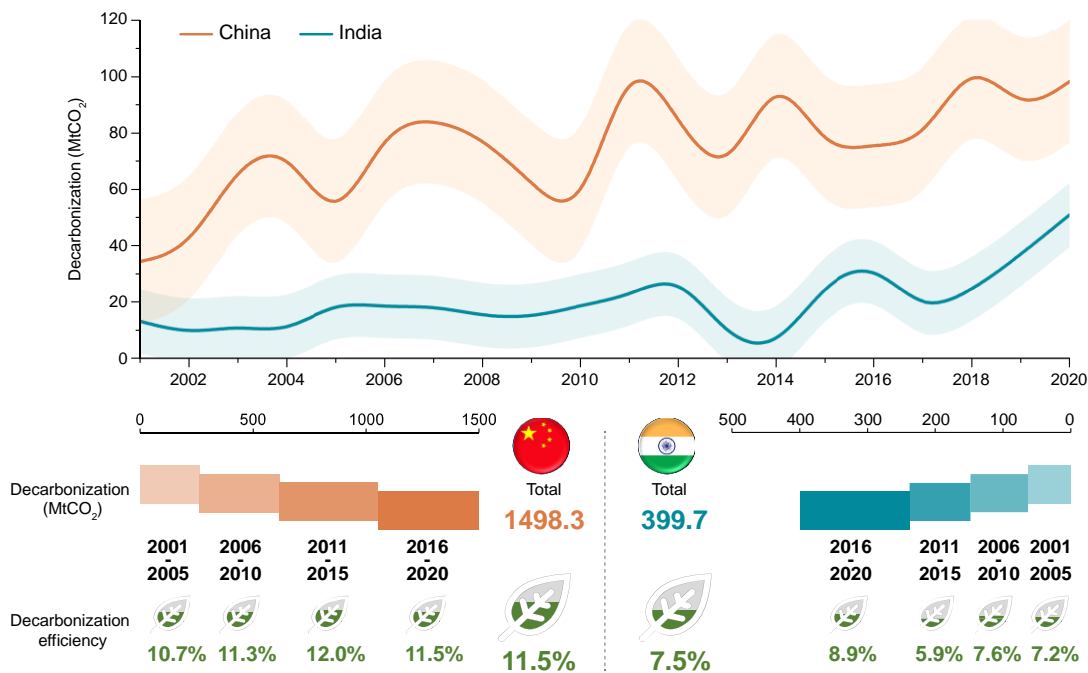

**Fig. 6.** The total decarbonization and decarbonization efficiency of residential building operations in India and China in 2000-2020.

In addition, this work aimed at investigating the decarbonization of residential building operations in India and China from the perspective of decarbonization intensity, per capita, per floor area, and per household expenditure. See Fig. 7 a: the decarbonization intensity in China averaged 167.3 kgCO$_2$ per household per year, with a growth of 3.7%/year. The yearly decarbonization intensity reached its peak in 2011 at 255.2 kgCO$_2$ per household, after which it decreased gradually and stabilized at a level above 150.0 kgCO$_2$ per household. The accumulated share between 2015 and 2020 in the total cumulative decarbonization intensity still amounted to 26%. In contrast, the decarbonization intensity in India has grown rapidly, particularly since 2015, with the accumulated share between 2015 and 2020 in the total cumulative decarbonization intensity reaching 36%. India's annual decarbonization intensity in 2020 reached 164.8 kgCO$_2$ per household, nearly catching up with that of China (182.5 kgCO$_2$ per household). Overall, although the decarbonization intensity in India was lower than that in China from 2000 to 2020, India is expected to surpass China in the upcoming years due to its robust growth of 7.3% per year.

Both China and India are the most populous countries, with each having a population of approximately 1.4 billion. However, when it comes to decarbonization per capita (as shown in Fig. 7 b), China has significantly outpaced India. Over the past two decades, China's per capita decarbonization has continued to rise, with an annual average of 55.5 kgCO$_2$ per capita and an increase of 5.1% per year. In contrast, India's per capita decarbonization has averaged 16.5 kgCO$_2$ annually, with slightly higher growth of 5.8% per year. Between 2015 and 2020, the share of the accumulation in the total cumulative decarbonization per capita reached 37%, coinciding with the rapid development of decarbonization efforts in India.

Although there are yearly fluctuations, the historical annual decarbonization per floor area in India and China are relatively similar, as demonstrated in Fig. 7 c. In China, the average decarbonization per floor area was 1.7 kgCO$_2$ per square meters (m$^2$) per year, with an increase of 1.9%/year. The peak of annual decarbonization per floor area occurred in 2011 at 2.6 kgCO$_2$/m$^2$, followed by a gradual decline. In India, the average decarbonization per floor area was 1.5 kgCO$_2$/m$^2$/year, with an increase of 3.0%/year. Notably, the decarbonization per floor area in India reached its highest value of 2.7 kgCO$_2$/m$^2$ in 2020 (compared to1.6 kgCO$_2$/m$^2$ in China), and is expected to continue increasing significantly.

Additionally, the historical performance of decarbonization per household expenditure for residential building operations is depicted in Fig. 7 d. In China, the decarbonization per household expenditure has experienced a sharp decline since 2004, with an average annual reduction of -4.3% over a twenty-year period. This reduction is mainly driven by the rapid growth in spending power and personal incomes in China, which outpaced the overall progress in decarbonization. In contrast, India saw only a slight decrease in decarbonization per household expenditure during the same twenty years, with a reduction of -0.01% per year averagely. The decarbonization per household expenditure values for China and India have converged since around 2015, reaching 10.7 kgCO$_2$ per thousand dollars and 9.0 kgCO$_2$ per thousand dollars in 2020, respectively. Similar to the differences in decarbonization intensity between the two countries, a reversal can also be anticipated in the coming years.

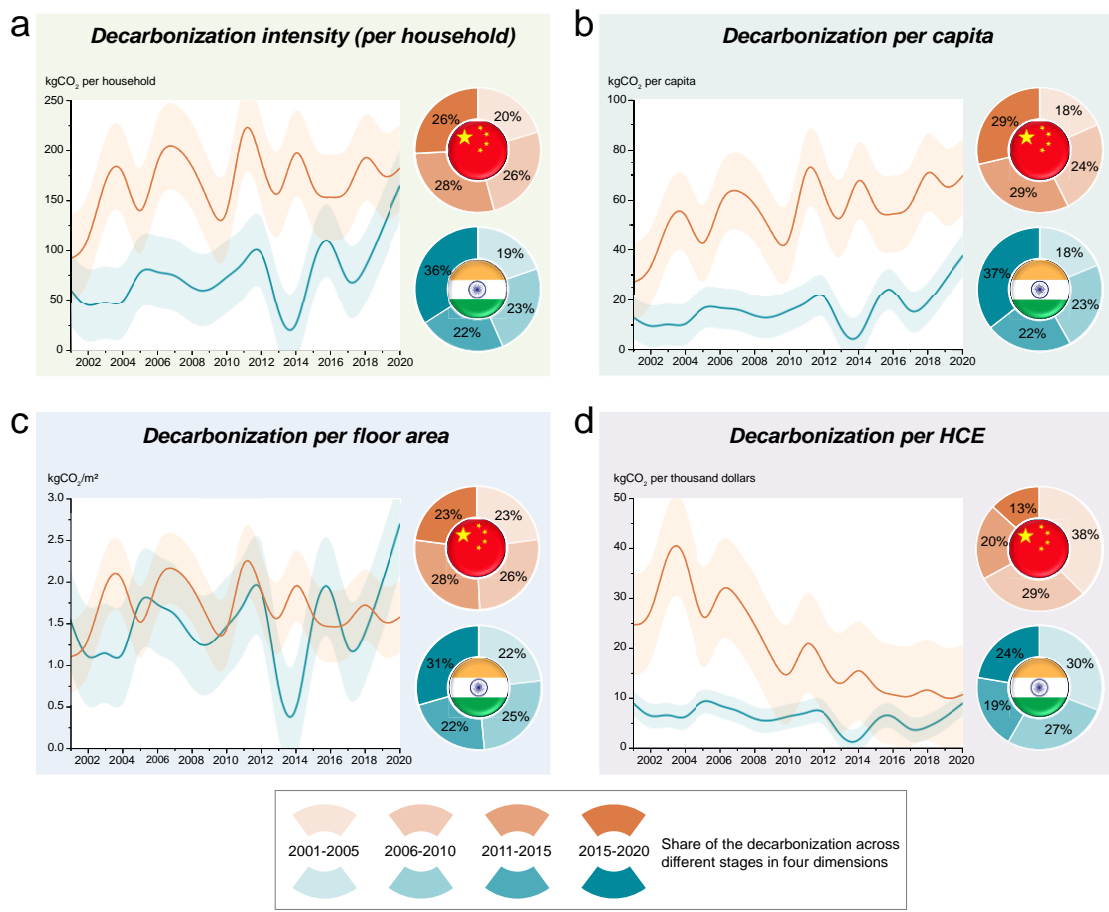

**Fig. 7.** The historical performance of (a) decarbonization per household, (b) per capita, (c) per floor area, and (d) per household expenditure of residential building operations in India and China during 2000-2020.

To summarize, the historical decarbonization performances of residential building in India and China were assessed comprehensively by the six scales of decarbonization mentioned above, completing the answer to Question 1 outlined in Section 1.

## 5.2. Robustness of the DSD-based decomposition

This work employed the DSD method to perform a three-layer decomposition of the carbon intensity model associated with end-use activities. While the DSD method has been utilized to assess the decarbonization of commercial building and effectively verified the robustness of the first layer decomposition of the emission model [44], the robustness analyses for the second layer decomposition of emission factor effects of different end uses, as well as the third layer decomposition of structural change effects of different end uses have not yet been demonstrated. On the other hand, LMDI is a typical decomposition approach and has been applied in carbon emissions research widely. However, due to the interdependence of factors, LMDI can only be expanded to a two-layer decomposition at most, and it cannot accurately decompose the structural change effects of different end uses. Therefore, this study introduced a two-layer LMDI approach to verify the robustness of the first and the second decomposition results calculated using the DSD method. Regarding the third layer decomposition of structural change effects of different end uses, verification is not considered here, as Section 4.2 already demonstrated that the structural change effects contributed minimally to carbon intensity changes.

According to the principles of the LMDI method and with consideration for factor consistency, the emission model of residential building in Section 3.1 [see Eq. (4)] was overwritten as:

$$c = \sum_{i=1}^{6} e_i \cdot k_i \cdot p \cdot g \cdot s = e \cdot k \cdot wi \cdot p \cdot g \cdot s \tag{20}$$

Referring to existing research on the two-layer LMDI method [57], the first layer intensity decomposition of carbon changes in residential building can be expressed as follows:

$$\Delta c|_{0 \to T} = \Delta p_{\text{LMDI}} + \Delta g_{\text{LMDI}} + \Delta s_{\text{LMDI}} + \Delta e_{\text{LMDI}} + \Delta k_{\text{LMDI}} \tag{21}$$

Using factor $p$ as an example, the expression of each factor's effect is as follows:

$$\Delta p_{\text{LMDI}} = \frac{c|_T - c|_{T-1}}{\ln c|_T - \ln c|_{T-1}} \cdot \ln\left(\frac{p|_T}{p|_{T-1}}\right) \tag{22}$$

The effect of factor $k$ can be further decomposed across six end uses, and the second layer decomposition can be obtained as follows:

$$\Delta k_{\text{LMDI}} = \sum_{i=1}^{6} \Delta k_{i_{\text{LMDI}}} = \sum_{i=1}^{6} \frac{c_i|_T - c_i|_{T-1}}{\ln c_i|_T - \ln c_i|_{T-1}} \cdot \ln\left(\frac{k_i|_T}{k_i|_{T-1}}\right) \tag{23}$$

Taking every five years as intervals, regression analysis was conducted for the comparison of the two-layer decomposition results by the DSD and LMDI methods, as shown in Fig. 8. All factors affecting carbon intensity changes were located in quadrants I and III, indicating that the factors' contributions resulting from the two decomposition methods were consistently positive or negative. All of the goodness-of-fit ($R^2$) values approached 1, and the length of the 95% confidence interval was close to 0, indicating that the two-layer decomposition results calculated by the DSD and LMDI methods are approximately equal.

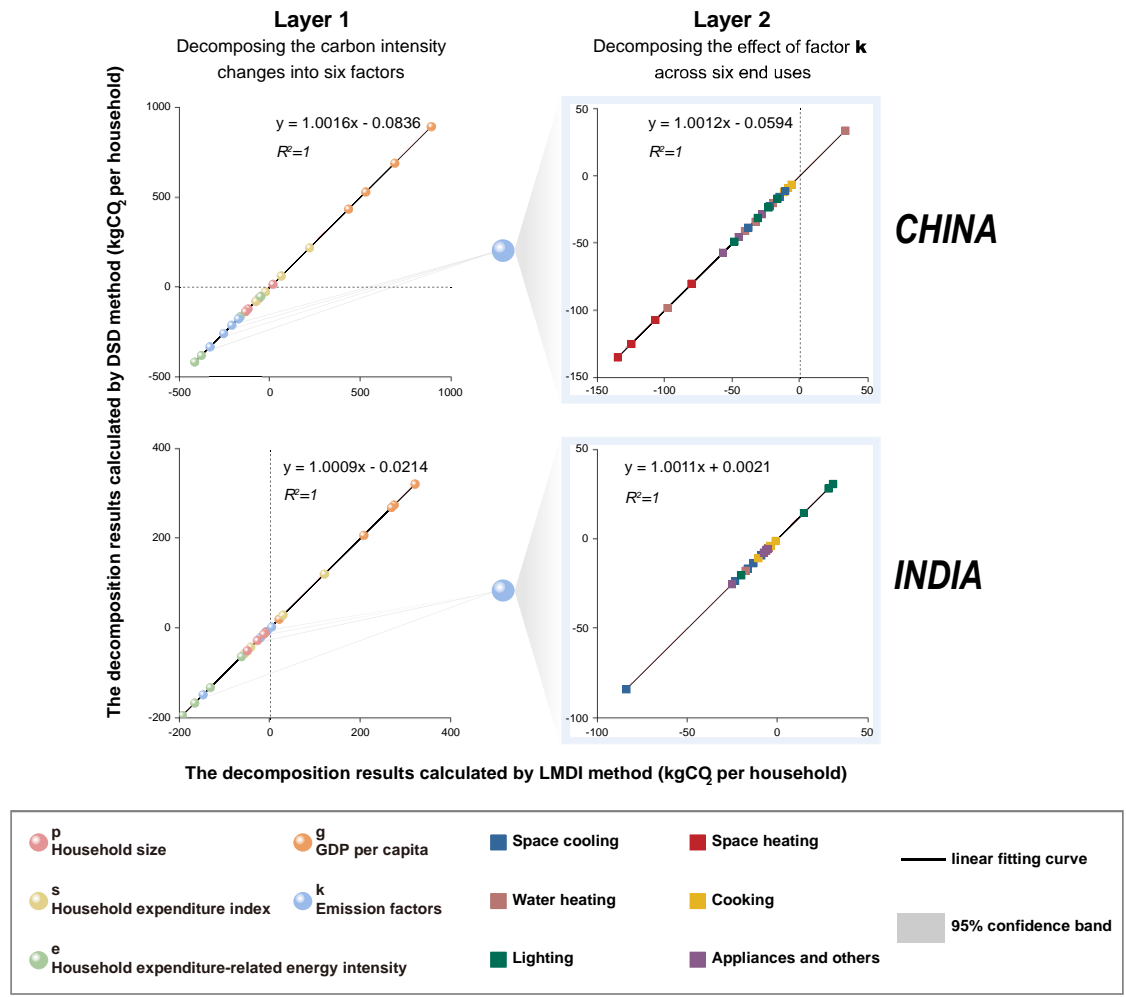

**Fig. 8.** Regression-based comparison of two-layer decomposition results by DSD and LMDI method.

Additionally, this study compared its results with previous studies to further verify their validity. Due to limited space, these comparisons are detailed in Appendix C. Overall, the DSD-based decomposition of the carbon intensity of residential building with end-use characteristics has been shown to be reliable. This finding strengthens the answers and arguments responding to Questions 1 and 2 raised in Section 1.

*5.3. Review and outlook on decarbonization strategies of residential building*

This work presented an overview of current policies and implementation strategies related to energy efficiency and decarbonization of residential building in India and China. It also provided recommendations for addressing existing challenges and achieving deep decarbonization in response to Question 3 outlined in Section 1. In the recent years, China and India have successively announced their respective goal of achieving carbon neutrality by 2060 and 2070 [58]. These announcements highlight not only the existing gap between the two countries in terms of decarbonization efforts, but also the potential for future competition and collaboration [59].

From the standpoint of current policies related to energy efficiency and decarbonization of residential building (see Fig. 9), China remains a step ahead of India in in terms of being more advanced, explicit, and systematic. China enacted the National Energy Conservation Law in the late 20th century and implemented the Regulation on Energy Conservation of Civil Buildings as early as 2005 [60]. Based on five-year plans, national strategies for building energy conservation have been consistently developed since the 1990s. Over the past decade, efficiency plans and schemes related to residential building operations have been proposed more frequently and with greater intensity, with the focus gradually shifting from energy conservation to decarbonization. Most notably, China's building energy efficiency standards are well established and comprehensive. Beginning with the initial Design Standard for Energy Efficiency of Residential Buildings adopted in 1986 (JGJ 26-1986), the system has progressively evolved by setting energy efficiency targets of 30%, 50%, 65%, and 75%. In 2019, the Technical Standard for Nearly Zero-Energy Building (GB/T 51350-2019) was introduced, which stipulated the energy performance of ultra-low energy, nearly zero-energy, and zero-energy buildings.

In contrast to China, India enacted the Energy Conservation Act in 2002. However, India lags behind China by 20 years when it comes to initiating building energy efficiency policies. The first energy efficiency standard, the Energy Conservation Building Code, was introduced in 2007 and amended in 2017. It primarily applied to large commercial buildings and is mostly a voluntary code. Thus far, the principle of voluntariness has been predominantly observed in India's relevant policies. It wasn't until 2018 that the energy efficiency standard for residential buildings, Eco Niwas Samhita, was launched in stages. This standard achieved energy savings of 20-25% compared to typical buildings. Similar to the Energy Conservation Building Code for commercial buildings, the implementation of Eco Niwas Samhita largely depends on state and local governments adopting and incentivizing the code, without strict enforcement from the national government [61]. Unlike Chinese local governments, which generally mandate the implementation of green building evaluation standards, India's green building rating and certification programs, such as the Indian Green Building Council or Green Rating for Integrated Habitat Assessment, are primarily voluntary and incentive-based policies [62]. Furthermore, the Indian government currently focuses on promoting plans and schemes related to end-use energy efficiency, while lacking energy efficiency strategies specifically for architectures.

**CHINA**      **INDIA**

**Laws and regulations**

CHINA:
- Interim Regulations on Energy Conservation (1986)
- National Energy Conservation Law of China(1997, 2007, 2016)
- National Renewable Energy Law of China (2015)

- Act on Energy Conservation of Civil Buildings (2005)

- China Energy Label Law (2004)
- Management Regulation for Certification of Energy-saving and Low-carbon Products (2015)

INDIA:
- Energy Conservation Act (2001)
- National Renewable Energy Act (2015)

**Plans and schemes**

CHINA:
- Medium and Long-term Plan of Energy Conservation (2004)
- Strategic Action Plan for Energy Development (2014)
- Energy Efficiency Leader Scheme (2016)
- Energy Supply and Consumption Revolution Strategy (2017)
- Three-year Action Plan for Cleaner Air (2018)
- Action Plan for Reaching Carbon Dioxide Peak Before 2030 (2018)

- 9th Five-Year-Plan (FYP) Building Energy Conservation Plan (1994)
- 10th FYP Building Energy Conservation Plan (1999)
- 12th FYP Building Energy Conservation Special Plan (2012)
- Green Building Action Plan (2013)
- 13th FYP for the Development of Building Efficiency and Green Building (2017)
- 14th FYP Comprehensive Work Plan for Energy Conservation and Emission Reduction (2022)

- Brightness Program (1997)
- Subsidies for Efficient Household Appliances (2012)
- Clean Winter Heating Plan in Northern China (2017)
- Green and High-Efficiency Cooling Action Plan (2019)

INDIA:
- Environment Action Program (1993)
- National Environment Policy (2006)
- Integrated Energy Policy (2006)
- National Action Plan on Climate Change (2008) including National Mission for Enhanced Energy Efficiency
- National Energy Efficient Buildings Program (2017)
- Draft National Energy Policy (2018)

- Energy Efficiency Label for Residential Sector (2019)

- Standards and Labelling Scheme(2006)
- Bachat Lamp Yojana (BLY) Lighting Program (2009)
- Super Efficient Equipment Program (2012)
- Unnat Jyoti by Affordable LEDs for All (2015)
- Pradhan Mantri Ujjwala Yojana Scheme (2016)
- Saubhagya Scheme (2017)
- India Cooling Action Plan (2019)

**Standards and guidelines**

CHINA:

Design & construction standards
Sorted by energy efficiency target (based on 1980s):
- 30%: JGJ 26-1986
- 50%: JGJ 26-1995, JGJ 134-2001, JGJ 75-2003, JGJ 134-2010, JGJ 475-2019
- 65%: JGJ 26-2010, JGJ 75-2012, GB/T 50824-2013
- 75%: JGJ 26-2018, GB 55015-2021
- ≥85%: GB/T 51350-2019

Evaluation & measurement standards
- JGJ/T 132-2009, GB/T 51366-2019
- GB/T 51141-2015, GB/T 50378-2019
- GB/T 51161-2016

AC:
- GB 12021.3-2010, GB 19576-2019
- GB 19577-2015, GB 21455-2019
- GB 35971-2018, GB 21454-2021

Heat pumps:
- GB 30721-2014, GB 37480-2019

Water heater:
- GB 21519-2008, GB 20665-2016
- GB 29541-2013

Cooker:
- GB 21456-2014, GB 30720-2014

Lighting:
- GB 30255-2019

Washing machine:
- GB 12021.4-2013

Refrigerator:
- GB12021.2-2015

INDIA:

Design & construction standards
Sorted by energy conservation target compared with typical buildings:
- 20-25%: Eco Niwas Samhita 2018, 2021

Evaluation & measurement standards
- Indian Green Building Council Green Ratings (2001)
- Green Rating for Integrated Habitat Assessment (2007)

AC:
- IS1391
Ceiling fans:
- IS 374:2019
Water heater:
- IS 2082:1993, IS 16368:2015
- IS 302-2-21:2011, IS 16543:2016
Cooker:
- IS 4246:2002
Lighting:
- IS 16102, IS 2418
Washing machine:
- IEC 60456, IS 302-2-7
Refrigerator:
- IS 1476:2000, IS 16590
- IS 15750:2006
TV:
- IEC 62301:2011, IS 616: 2017
- IEC 62087:2015

Legend: policies of integrated energy conservation | policies focused on buildings | policies focused on end-use facilities

**Fig. 9.** Review on decarbonization strategies of residential building in India and China.

In regards to the implementation of pertinent policies, it is valuable to examine the current progress and potential influence of residential electrification, as well as the renovation and new construction of energy-efficient residential buildings. These two strategies are increasingly being recognized as appealing methods for achieving decarbonization of residential building operations, through active and passive means, respectively [63].

In terms of active means, prioritizing the electrification of end-use energy consumption is crucial for deploying zero-carbon electricity [64]. By 2019, nearly all households in China and India had access to electricity through China's Brightness Program [65] and India's Saubhagya Scheme [66]. However, achieving comprehensive full electrification and decarbonizing electricity remains ongoing challenges. Fig. 10 illustrates the proportion of thermal power, the emission factor of electricity, and the electrification rate in residential building in India and China. Currently, the electrification rates in residential building operations are 27.2% for China and 17.8% for India. End-use activities such as cooking and heating still heavily rely on solid fuels [67], particularly coal and wood in rural areas. Electricity generation in China is still predominantly based on thermal power (69.5% in 2020 [68]), with coal being the primary source. The same holds true for India, where thermal power accounts for 75.2% of electricity generation in 2020 [69]).

To estimate the decarbonization potential achievable solely through electrification improvements by 2030, we assumed, based on recent studies, that the proportion of thermal power in China will decrease to 60% in 2030, the emission factor of electricity will drop to approximately 0.46 kgCO$_2$/kWh (i.e., 3.7 kgCO$_2$/kgce), and the electrification rate will reach 38% [70]. With Chinese households projected to reach 545 million in 2030 [71], and energy consumption per household continuing to increase at a historical rate of 4.8% per year. an additional 81.9 MtCO$_2$ would be cut in China if the emission factor of primary energy remains unchanged. Similarly, India is predicted to reach the China's 2020 electrification rate by 2030, with the proportion of thermal power decreasing to 70%, the emission factor of electricity dropping to approximately 0.56 kgCO$_2$/kWh (i.e., 4.5 kgCO$_2$/kgce), and the electrification rate will reach 28% [72]. With Indian households projected to reach 350 million in 2030 [73], and energy consumption per household continuing to increase at a historical rate of 2.3% per year, an additional 49.6 MtCO$_2$ would be reduced in India due to electrification improvements.

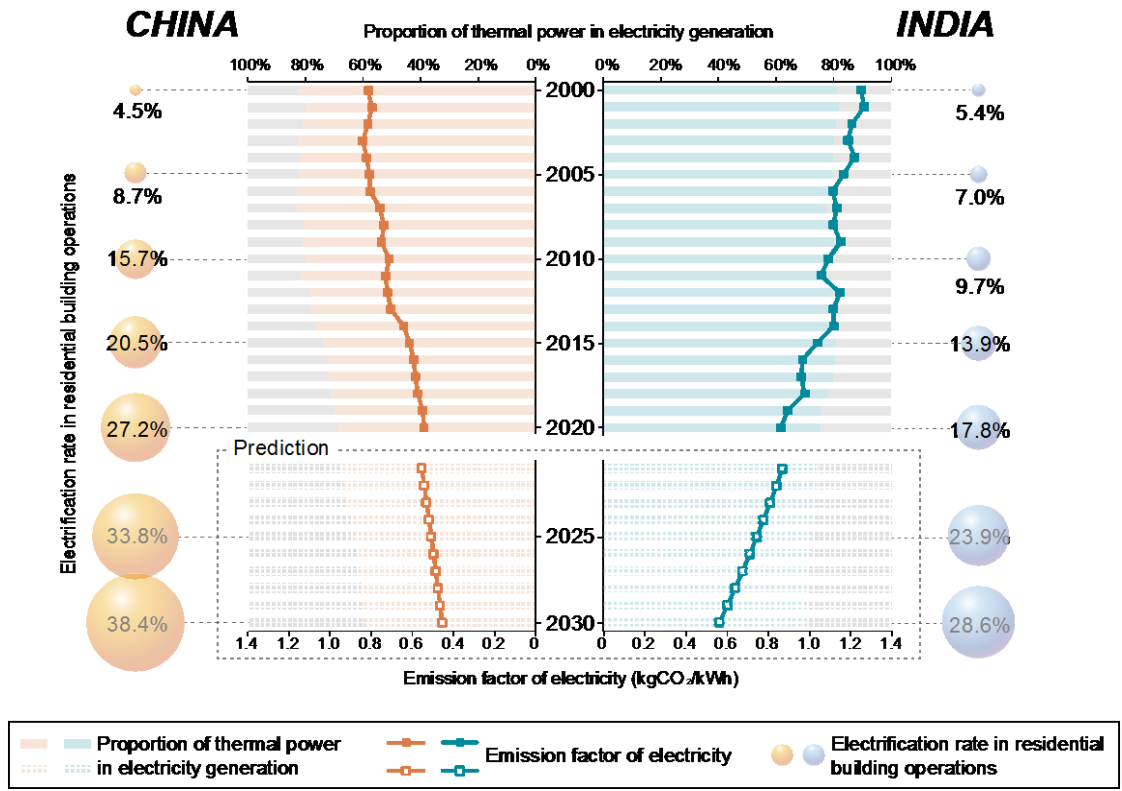

**Fig. 10.** Comparison of the proportion of thermal power, the emission factor of electricity, and the electrification rate in residential building operations between China and India during 2000-2030.

In terms of passive means, the renovation and new construction of energy-efficient buildings represent a crucial strategic transformation for India and China, the global two most populous economies, in their ongoing urbanization process. As previously mentioned, China has been developing a series of policies for around 40 years, establishing phased plans for the renovation and new construction of energy-efficient buildings in various climate zones. Between 2016 and 2020, 514 million $m^2$ of existing residential buildings were renovated for energy efficiency, resulting in a decarbonization of 14.3 MtCO$_2$. Additionally, approximately 10.5 billion $m^2$ of new energy-efficient buildings were constructed in urban areas, including 10 million $m^2$ of ultra-low energy buildings and nearly zero-energy buildings [74]. In line with China's target of achieving carbon neutrality by 2060, the renovation and new construction of residential energy-efficient buildings are expected to accomplish a cumulative decarbonization of 429 MtCO$_2$ (compared to China's current building energy efficiency standards) between 2020 and 2060, with decarbonization anticipated to reach 111 MtCO$_2$ by 2030 (see Appendix B for detailed targets) [75].

In contrast, while India is lagging behind, its government and relevant institutions have been actively promoting the Eco-Niwas Samhita, introduced in 2018. A comprehensive design plan adhering to the Eco-Niwas Samhita has been formulated for all five climate zones in India [76]. It is suggested that the new construction of affordable housing, which accounts for 95% of urban housing demand, fully implement the Eco Niwas Samhita [77], and that at least 5% of the existing residential building stock also undergo Eco-Niwas Samhita renovation [78]. By 2030, the implementation of this code is projected to save 125 billion kWh of electricity per year, equivalent to a decarbonization of approximately 100 MtCO$_2$ (compared to current non-energy-efficient buildings) [79]. The potential for further residential building decarbonization in India is immense.

According to the analysis above, the following proposals are suggested for implementing further strategies:

a. The electrification of end-use performances in residential building is the most effective mean for fully utilizing waste heat from industries, power plants, and sewage for space heating and cooling [80] as a medium-term plan. Furthermore, as electric vehicles become more popular, charging stations can be integrated into building energy systems to enable distributed energy storage [81]. A long-term pathway to complete electrification will involve the combination of electric heat pumps, renewable energy sources, and integrated energy storage [82].

b. Energy efficiency standards for new buildings are supposed to be enforced nationwide, and the design, construction and purchase of ultra-low energy buildings or zero-energy buildings can be promoted through the provision of subsidies [83]. Additionally, there should be an increased focus on energy-efficient renovations of existing building stock (e.g., Assessment Standard for Green Retrofitting of Existing Building, AKA GB/T 51141-2015 in China) to overcome the carbon lock-in, particularly in existing residential buildings [84]. Overall, this work calculates and compares the past process and current pattern of operational decarbonization in residential building of India and China, and it offers corresponding adaptive strategies to better lead future residential buildings of the emerging economies to the carbon neutral status after the mid-century, thereby addressing Question 3 outlined in Section 1.

## 6. Conclusion

This work evaluated the progress of operational decarbonization in residential building in India and China during the 21st century using the improved DSD method. First, a carbon intensity model associated with end-use performances was developed to identify the factors affecting the carbon changes, and the impacts of various end uses on the carbon intensity were further investigated. Moreover, six scales of decarbonization were employed to track the historical processes of decarbonizing residential buildings. Finally, a review and outlook on decarbonization strategies for residential building operations in India and China were developed to address current challenges and achieve significant decarbonization in the future. The main findings are briefly provided below.

### 6.1. Main findings

- **Operational carbon intensity increased at 1.4% (from 1125 to 1492 kgCO$_2$ per household) and 2.5% (from 744 to 1216 kgCO$_2$ per household) per year in China and India from 2000 to 2020, respectively.** Chinese residential buildings' carbon intensity reached an annual peak in 2012 with 1606.4 kgCO$_2$ per household, while that of India reached an annual peak at 1283.3 kgCO$_2$ per household in 2018. Moreover, GDP per capita was the most significant positive contributor, with a total contribution of 226.4% in China and 144.5% in India during 2000-2020. Household expenditure-related energy intensity was the most significant negative contributor (-90.6% in China and -78.1% in India), followed by emission factors (-87.4% in China and -23.0% in India), both of which were crucial to decarbonizing residential building operations.

- **Building electrification promoted the end-uses' emission factor effects on decarbonization (e.g., space cooling contributed -87.7 and -130.2 kgCO2 per household in China and India, respectively).** The most significant positive contributor to decarbonization in China was space heating, with a total contribution of -39.8% and an overall impact on carbon intensity changes of -447.9 kgCO$_2$ per household from 2000 to 2020. This was followed by appliances and others with -15.1% (-169.7 kgCO$_2$ per household), and lighting contributed -10.7% (-120.6 kgCO$_2$ per household). In India, the most significant positive contributor to decarbonization was space cooling, followed by appliances and others, with -5.8% (-43.4 kgCO$_2$ per household). Although lighting turned into a positive effect on decarbonization from 2015, it remained the most significant negative contributor (7.6%, 56.4 kgCO$_2$ per household).

- **China and India collectively decarbonized 1498.3 and 399.7 MtCO$_2$ of residential building operations from 2000 to 2020, but neither reached an annual peak of decarbonization.** China led in decarbonization efficiency at 11.5%, compared to India's 7.5%. Additionally, the decarbonization efficiency in the United States was 8.5% over the same period, indicating an inverted-U shaped relationship between decarbonization efficiency and national development. However, India's operational decarbonization intensity is likely to surpass China's in the next few years, as India's annual decarbonization intensity reached 164.8 kgCO$_2$ per household in 2020, close to China's 182.5 kgCO$_2$ per household. In recent years, the decarbonization per floor area and decarbonization per household expenditure have been relatively similar between China and India.

*6.2. Upcoming works*

To explore the best practical pathways for high decarbonization in residential building sector of the two largest emerging emitters, the following gaps can be addressed in future work: (1) Extend the research perspective to future decarbonization trend. Assess the achievability of national decarbonization targets under scenarios of business-as-usual, nationally determined contributions, and global warming of 1.5-2 °C to seek the optimal path for carbon neutrality. (2) Review and evaluate the policy implementation of energy efficiency and decarbonization initiatives and regulations in the world's major emitters through policy informatics, making recommendations for public policy design of residential building decarbonization strategies in both China and India. (3) Analyze decarbonization methods of residential buildings with smart urban governance models to address differences in climate, population, and economy among various regions. Investigate the cooperative relationships among residents, related enterprises, and local governments to contribute to a more targeted deployment of decarbonization strategies.

## Acknowledgment

This manuscript has been authored by an author at Lawrence Berkeley National Laboratory under Contract No. DE-AC02-05CH11231 with the U.S. Department of Energy. The U.S. Government retains, and the publisher, by accepting the article for publication, acknowledges, that the U.S. Government retains a non-exclusive, paid-up, irrevocable, world-wide license to publish or reproduce the published form of this manuscript, or allow others to do so, for U.S. Government purposes.